\documentclass[twocolumn]{cinc}
\usepackage{graphicx}
\usepackage{multirow}
\usepackage{booktabs}
\usepackage{makecell}
\usepackage{graphicx}
\usepackage{amsmath,amsfonts,amssymb}
\begin{document}
\bibliographystyle{cinc}

% Keep the title short enough to fit on a single line if possible.
% Don't end it with a full stop (period).  Don't use ALL CAPS.
\title{Dynamic Heartbeat Modeling with Recurrent Neural Networks and Inverse Gaussian Point Process}

% Both authors and affiliations go in the \author{ ... } block.
% List initials and surnames of authors, no full stops (periods),
%  titles, or degrees.
% Don't use ALL CAPS, and don't use ``and'' before the name of the
%  last author.
% Leave an empty line between authors and affiliations.
% List affiliations, city, [state or province,] country only
%  (no street addresses or postcodes).
% If there are multiple affiliations, use superscript numerals to associate
%  each author with his or her affiliations, as in the example below.

\author {Runwei Lin, Ying Wang \\
Biomedical Signals and Systems, University of Twente, Enschede, the Netherlands}

\maketitle

% LaTeX inserts the ``Abstract'' heading in the proper style and
% sets the text of the abstract in italics as required.
\begin{abstract}
Heart rate variability (HRV) analysis is important for the assessment of autonomic cardiovascular regulation. The inverse Gaussian process (IGP) has been widely used for beat-to-beat HRV modeling. However, dynamic estimation of the IGP parameters remains a critical challenge. Conventional parametric approaches such as autoregressive models might be insufficient to capture the nonlinear characteristics of R-R series. In this study, we investigated whether recurrent neural networks (RNNs) can be used for IGP parameter identification and thereby enhance probabilistic modeling of R-R dynamics. Four representative RNN architectures, namely, GRU, LSTM, Structured State Space sequence model (S4), and Mamba, were compared and evaluated using the Kolmogorov-Smirnov statistics. The results demonstrate the possibility of combining RNN with the IGP for beat-wise R-R series modeling. This approach provides a flexible basis for probabilistic HRV modeling and for future incorporation of more complex physiological mechanisms and dynamic conditions.
\end{abstract}

\section{Introduction}
Modeling of R-R dynamics at the beat-level is essential in computational cardiology. Variations in heartbeat, commonly known as heart rate variability (HRV), reflect autonomic cardiac regulation and are clinically relevant for the assessment of autonomic dysfunction, arrhythmic risk, and cardiovascular aging \cite{acharya2006heart,shaffer2017overview}.
Traditional HRV indices are typically calculated over windows ranging from minutes to hours \cite{shaffer2017overview}. 
Although such features are useful in many applications, they compress R-R series into aggregate statistical measures and therefore provide limited expression of the sequential structure, which can be especially important under nonstationary physiological conditions. Probabilistic formulations that enable beat-level modeling and uncertainty quantification are therefore important in such cases.

Point processes are attractive in characterizing beat-to-beat R-R dynamics, as they treat each heartbeat as a stochastic event through a conditional probability framework \cite{barbieri2005point}. 
A widely used point-process formulation for heartbeat modeling is the inverse Gaussian process (IGP), in which the occurrence of the next R peak is modeled as the first-passage time of a Brownian motion with linear drift \cite{barbieri2005point}. This interpretation is physiologically appealing as it is consistent with the integrate-to-threshold process of sinoatrial node activation \cite{barbieri2005point}. 

Parameter estimation is the central challenge for IGP based R-R modeling.
Conventional IGP models typically use parametric models to characterize the evolution of heartbeat dynamics \cite{barbieri_analysis_2006,chen_characterizing_2010,citi2012real,valenzaPointProcessNonlinearModels2013a}. However, HRV is inherently nonlinear and exhibits substantial inter-individual variability. As a result, these parametric forms may be limited in fully capturing the complex HRV dynamics. Recent neural temporal point process methods have introduced more flexible heartbeat modeling methods \cite{subramanian2025density}, yet their focus is on density modeling to enhance model accuracy. This leaves the use of recurrent neural networks (RNNs) for adaptive IGP parameter inference largely unexplored.

In this work, we integrate RNNs with IGP for beat-to-beat heartbeat modeling. Specifically, GRU, LSTM, Structured State Space sequence model (S4), and Mamba were used as representative architectures to learn the dynamic evolution of IGP parameters directly from heartbeat history \cite{lecun2015deep, gu2021efficiently, gu2024mamba}. This design combines neural networks with the physiological interpretability of IGP. The proposed framework provides a flexible alternative to conventional methods, and may serve as a basis for future analysis of dynamic autonomic cardiovascular regulation.

\section{Method}
\subsection{Data description and preprocessing}
This study uses the MIT-BIH Normal Sinus Rhythm Database for validation \cite{pollard2026physionet}. The dataset includes 18 healthy subjects (5 men, aged 26 to 45, and 13 women, aged 20 to 50) and has been used in previous studies as benchmark \cite{barbieri_analysis_2006,chen_characterizing_2010,subramanian2025density}. 
Long-term ECG were measured at 128Hz for 2-3 hours.
R peaks were identified by the Pan-Tompkins method \cite{pan1985real}, R-R intervals below 0.3s or above 2s were regarded as invalid and excluded from the recordings. The excluded values were then interpolated with the piecewise cubic Hermite interpolating polynomial method. 

\subsection{Inverse Gaussian process}
IGP models the heartbeat sequence in continuous time via the occurrence times of successive R peaks. Let $r_i$ denote the occurrence time of the $i$th R peak, and define
\begin{equation}
x_{i+1} = r_{i+1} - r_i
\end{equation}
as the next R--R interval, i.e., the waiting time from the current R peak to the next one. Given the heartbeat history up to beat $i$, denoted by ${x}_{1:i}$, the IGP assumes that the conditional distribution of $x_{i+1}$ follows an inverse Gaussian distribution:
\begin{equation}
p(x_{i+1}\mid {x}_{1:i})
=
\sqrt{\frac{\lambda_i}{2\pi x_{i+1}^3}}
\exp\left[
-\frac{\lambda_i (x_{i+1}-\mu_i)^2}{2\mu_i^2 x_{i+1}}
\right],
\end{equation}
where $\mu_i>0$ is the conditional mean of the next R--R interval, and $\lambda_i>0$ is the shape parameter. The conditional standard deviation is:
\begin{equation}
\sigma_i = \sqrt{\frac{\mu_i^3}{\lambda_i}}.
\label{sigma}
\end{equation}

\subsection{Network architecture}
\begin{figure}[!htb]
    \centering
    \includegraphics[width=\linewidth]{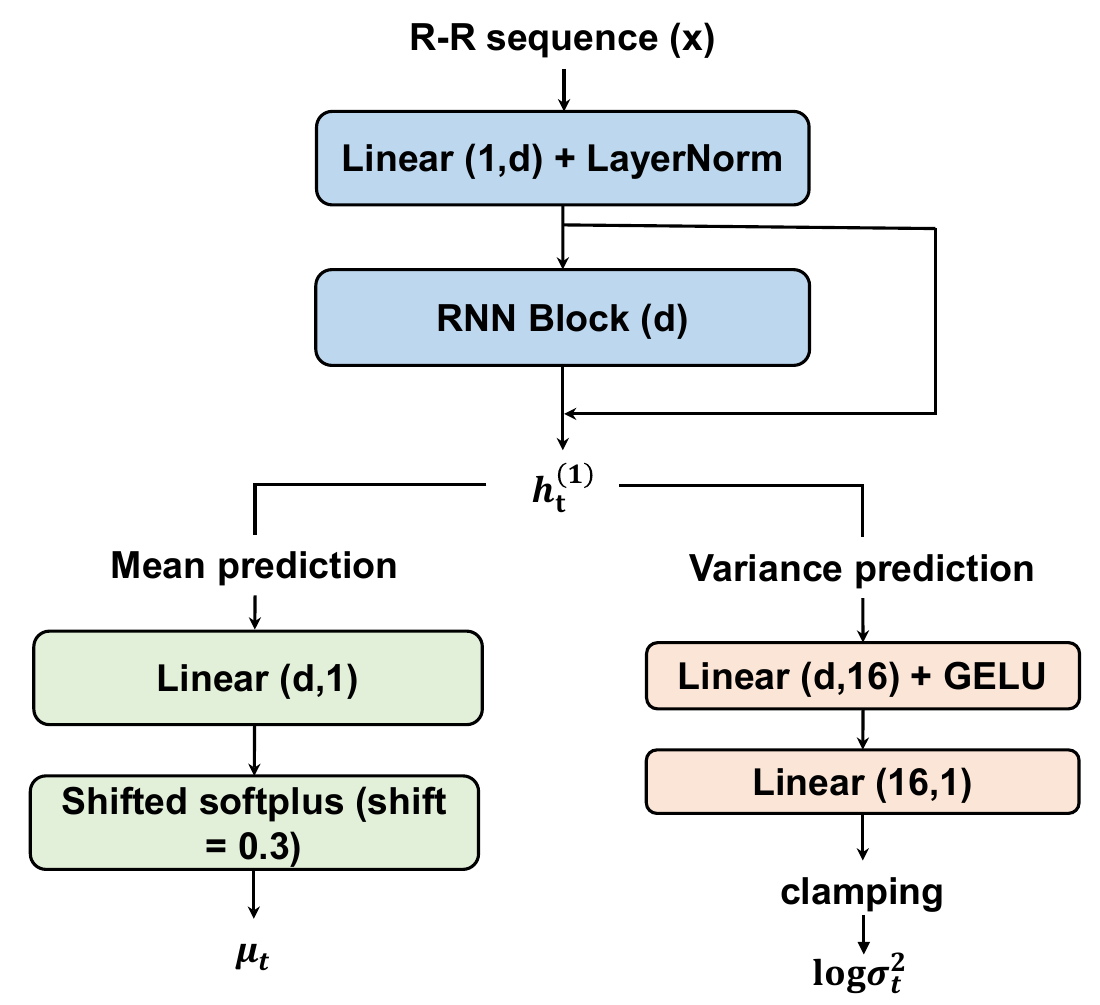}
    \caption{The R–R sequence is encoded by a sequence model backbone and mapped to time-varying inverse Gaussian parameters. Separate branches are used to estimate the mean and the log-variance.}
    \label{scheme}
\end{figure}

We modeled beat-to-beat R-R dynamics using a causal sequence architecture. Figure \ref{scheme} shows the network structure.
Let $\textbf{x} = [x_1,\dots,x_T]^\top$ denote the input R-R sequence, where $x_t$ is the R-R duration at $t$th R-R interval. 
Output of the neural network is the corresponding time-varying IGP parameters at each beat.
Each scalar input was first projected to a $d$-dimensional linear embedding with layer normalization. 
The embedded sequence was then processed by an RNN block $\mathcal{B}(\cdot)$ with residual connections:
\begin{align}
\mathbf{h}_t^{(1)} &= \mathbf{h}_t + \mathcal{B}(\mathbf{h}_{1:t}).
\end{align}
Two prediction branches were constructed from the hidden representation $\mathbf{h}_t^{(1)}$. First, a linear head produced a latent trend:
\begin{equation}
z_t = W_\mu \mathbf{h}_t^{(1)} + b_\mu,
\end{equation}
which was mapped to a strictly positive observation mean through a shifted softplus transform,
\begin{equation}
\mu_t = \text{softplus} (z_t) + \delta,
\end{equation}
where $\delta = 0.3$ is a small constant to ensure the predicted mean stays above 0.3s (below 200 bpm). Second, an uncertainty head predicted the log-variance,
\begin{equation}
\log \sigma_t^2 = f_\sigma(\mathbf{h}_t^{(1)}),
\end{equation}
where $f_\sigma(\cdot)$ was implemented as a two-layer MLP.
We chose to predict $\log \sigma_t^2$ instead of $\lambda_t$ as the uncertainty is more of interest in this study. In \cite{pessoaMambaTimeSeries2025}, authors suggested that sequence framework like Mamba are better at capturing deterministic evolution than uncertainty, and therefore an MLP can better capture the uncertainty variation. We thus chose a similar approach to predict the log-variance. The $\log \sigma_t^2$ was clipped to [-9,1.5] (empirical threshold) during training, which corresponds to a standard deviation of [11.1ms, 2.12s].
To enable a fair architectural comparison, $\mathcal{B}(\cdot)$ was constructed by comparing a single GRU, LSTM, S4 and Mamba layer. 

\subsection{Model training}
The IGP parameters were estimated by minimizing the total negative log-likelihood of the inverse Gaussian distribution over all input R--R intervals,
where the per-step negative log-likelihood is given by:
\begin{equation}
\ell_i =
\frac{1}{2}\log\!\left(\frac{2\pi x_{i+1}^3\sigma_i^2}{\mu^3_i}\right)
+
\frac{\mu_i (x_{i+1}-\mu_i)^2}{2\sigma_i^2 x_{i+1}}.
\end{equation}
where $x_{i+1}$ denotes the observed R--R interval, and $(\mu_i, \sigma_i)$ are parameters at step $i$ conditioned on previous heartbeat history. Accordingly, the loss function $L$ is:
\begin{equation}
    L = \sum\limits_{i=1}^T \ell_i
\end{equation}
$T$ is the length of the input sequence.

To evaluate cross-subject generalizability, we adopted a leave-one-subject-out cross-validation scheme. In each run, 16 subjects were used for training, one for validation, and one for testing, resulting in 18 runs such that each subject served once as the test subject. 
Considering the duration of each recording, R–R interval series were segmented into non-overlapping sequences to construct sufficient training samples and to enable evaluation under different temporal contexts. We considered four training sequence lengths (100, 200, 300, and 600 beats).
For testing, non-overlapping segments of 600 and 1800 beats were used to assess model performance over short- and long-term dynamics, respectively. 
All models were trained for up to 2000 epochs with early stopping. The model dimension was fixed at 64, and a single-layer architecture was used for all models. For Mamba, the state dimension was set to 32 based on preliminary comparisons (8, 16, 32, 64), and for S4, the state dimension was fixed at 32 for consistency.
\subsection{Performance Evaluation}
Model goodness-of-fit (GOF) was assessed using the Kolmogorov--Smirnov distance (KSD), following previous studies \cite{barbieri_analysis_2006,chen_characterizing_2010,valenzaPointProcessNonlinearModels2013a,subramanian2025density}. Specifically, for each observed interval $x_{i+1}$, the fitted conditional cumulative distribution function (CDF) was used to compute:
\begin{equation}
u_i = F_{\mathrm{IG}}\!\left(x_{i+1}\mid \mu_i,\sigma_i\right),
\end{equation}
where $F_{\mathrm{IG}}(\cdot \mid \mu_i,\sigma_i)$ denotes the inverse Gaussian CDF with parameters predicted from the heartbeat history. According to the time-rescaling theorem \cite{brown2002time}, the transformed samples $\{u_i\}$ are independent and identically distributed as $U(0,1)$ if the model is correctly specified. 
The K-S plot was then constructed by comparing the empirical cumulative distribution of $\{u_i\}$ against the identity line $y=x$. The KSD was defined as the maximum vertical deviation between the empirical distribution and the identity line. For $T$ transformed samples, the 95\% confidence bound is $1.36/\sqrt{T}$, with smaller KSD indicating better model fit \cite{brown2002time,barbieri2005point,barbieri_analysis_2006,subramanian2025density}. In this study, the corresponding 95\% confidence bounds are 0.056 (600 beats) and 0.032 (1800 beats). 
\section{Results}
\begin{figure}[!htb]
    \centering
    \includegraphics[width=\linewidth]{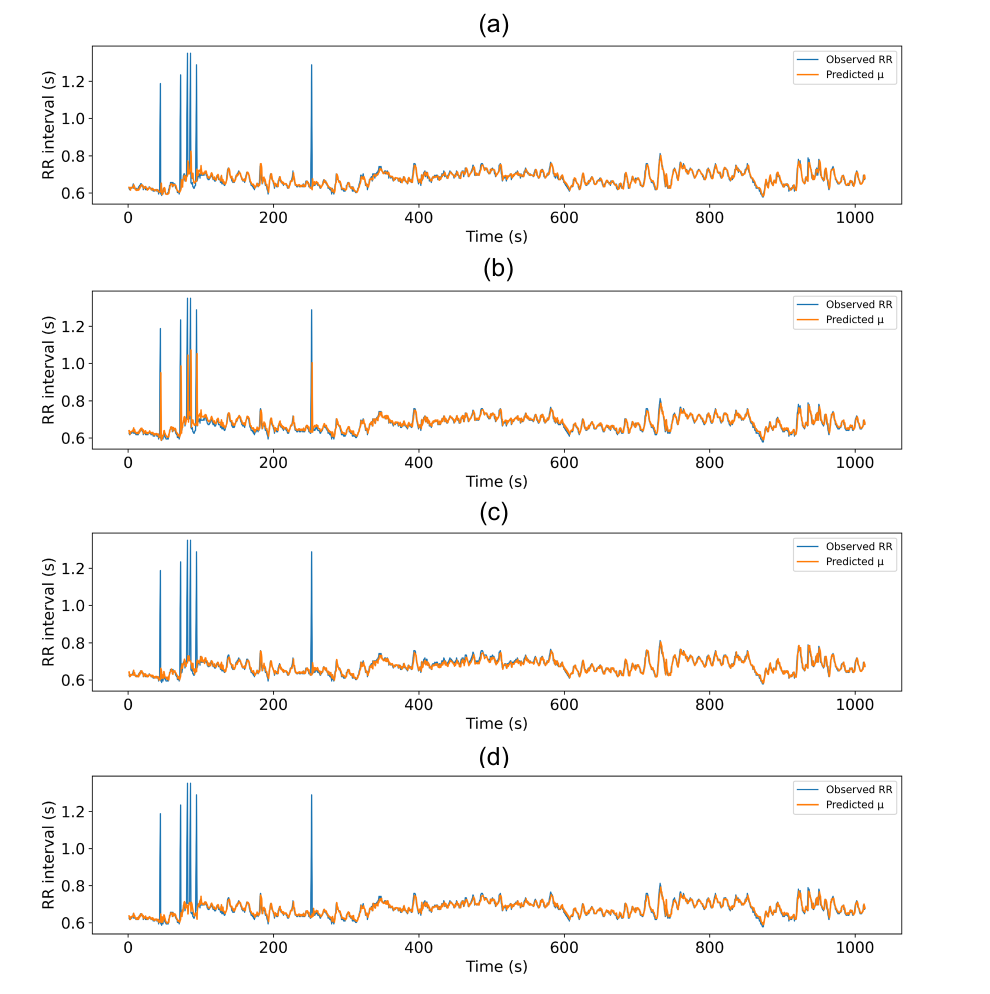}
    \caption{A representative example of the testing results whose length including 1800 beats from comparing the predicted $\mu_t$ and the measured R–R interval series from subject 16483. Blue: Measured R-R; Orange: $\mu_t$ prediction. (a) GRU, (b) LSTM, (c) S4, and (d) Mamba.}
    \label{res_eg}
\end{figure}

Our results showed that training length of 600 beats yielded the lowest mean KSD. 
Tables \ref{tab:600} and \ref{tab:subjectwise_ksd} report the mean and standard deviation of the test KSD for each subject under two testing segment lengths (600 and 1800 beats), both with training length of 600 beats. Figure \ref{res_eg} gives an example comparison between the R-R measurement and $\mu_t$ estimated by different types of RNNs. 

\begin{table}[h]
\centering
\scriptsize
\caption{Test results (Mean [SD]) of KS distance with the testing segment of 600 beats. The last row reports the across-subject mean of subject-level mean KSD.}

\begin{tabular}{cccccc}
\hline
\textbf{Subject ID} &\textbf{GRU} & \textbf{LSTM} &\textbf{S4} & \textbf{Mamba} \\
\hline
16265&  0.114 [0.056] & 0.124 [0.058] &0.244 [0.042]& 0.113 [0.060]  \\
16272  & 0.103 [0.056]  & 0.106 [0.055] & 0.136 [0.045] & 0.122 [0.054]  \\
16273  & 0.070 [0.030]& 0.077 [0.025] & 0.071 [0.031] & 0.060 [0.025]  \\
16420  & 0.074 [0.030]& 0.078 [0.030] & 0.089 [0.026] & 0.103 [0.039]  \\
16483 & 0.114 [0.040]& 0.178 [0.036] & 0.177 [0.027] & 0.107 [0.027] \\
16539  & 0.124 [0.025]& 0.086 [0.031] & 0.104 [0.023] & 0.114 [0.032] \\
16773 & 0.096 [0.038]& 0.116 [0.083] & 0.111 [0.044]& 0.086 [0.037]\\
16786 & 0.093 [0.050]& 0.086 [0.054] & 0.071 [0.035] & 0.081 [0.053] \\
16795 & 0.066 [0.026] & 0.063 [0.022] &0.068 [0.029]& 0.071 [0.028] \\
17052 & 0.070 [0.022]& 0.073 [0.027] & 0.087 [0.038] & 0.067 [0.026]\\
17453 & 0.103 [0.022]& 0.070 [0.024] & 0.092 [0.025]  & 0.082 [0.023] \\
18177 & 0.117 [0.043]& 0.124 [0.042] & 0.095 [0.034] & 0.115 [0.041] \\
18184 & 0.060 [0.019] & 0.095 [0.035] & 0.114 [0.032] &  0.087 [0.029] \\
19088 & 0.075 [0.033] & 0.073 [0.031] & 0.089 [0.029] &  0.082 [0.030] \\
19090 & 0.082 [0.027] & 0.062 [0.020] & 0.097 [0.034]  & 0.067 [0.025] \\
19093 & 0.062 [0.027] & 0.061 [0.023] & 0.074 [0.025] & 0.082 [0.027] \\
19140 & 0.068 [0.027] & 0.061 [0.021] & 0.077 [0.026]  & 0.070 [0.024] \\
19830 & 0.174 [0.069]& 0.179 [0.064] & 0.539 [0.053] & 0.177 [0.069] \\
\hline
Overall Mean& 0.0925  & 0.0951&  	0.1297&0.0937\\
\hline
\end{tabular}
\label{tab:600}
\end{table}

\begin{table}[h]
\centering
\scriptsize
\caption{Test results (Mean [SD]) of KS distance with the testing segment of 1800 beats. The last row reports the across-subject mean of subject-level mean KSD.}
\begin{tabular}{ccccc}
\hline
\textbf{Subject ID} &\textbf{GRU} & \textbf{LSTM} &\textbf{S4} & \textbf{Mamba} \\
\hline
16265  & 0.115 [0.057]  & 0.125 [0.060] & 0.202 [0.041] &0.102 [0.051] \\
16272  & 0.097 [0.050]  & 0.097 [0.051] & 0.230 [0.034] & 0.115 [0.056] \\
16273  & 0.053 [0.021] & 0.066 [0.016] & 0.107 [0.038] & 0.044 [0.018]  \\
16420  & 0.060 [0.023] & 0.062 [0.025] & 0.117 [0.038] & 0.087 [0.034]  \\
16483 & 0.108 [0.046] & 0.167 [0.030] & 0.129 [0.026] & 0.100 [0.024] \\
16539  & 0.127 [0.020] & 0.083 [0.025]& 0.117 [0.025] & 0.099 [0.024] \\
16773 & 0.098 [0.035]  & 0.107 [0.077] & 0.152 [0.087] & 0.081 [0.035] \\
16786 & 0.084 [0.052] & 0.084 [0.056] &0.078 [0.034] & 0.074 [0.058] \\
16795 & 0.053 [0.023] & 0.051 [0.019] & 0.060 [0.021]& 0.060 [0.027] \\
17052 & 0.063 [0.018] & 0.060 [0.026] & 0.203 [0.038]  & 0.062 [0.027] \\
17453 & 0.094 [0.016] & 0.059 [0.018] & 0.134 [0.043] & 0.075 [0.027]\\
18177 & 0.106 [0.039] & 0.115 [0.033] & 0.288 [0.052]  &0.109 [0.037] \\
18184 & 0.044 [0.018]&  0.084 [0.028] &  0.144 [0.043]  & 0.076 [0.021] \\
19088 & 0.065 [0.027] & 0.060 [0.024] & 0.105 [0.023] & 0.064 [0.026]  \\
19090 & 0.072 [0.023]& 0.046 [0.015] & 0.097 [0.033]  & 0.054 [0.022] \\
19093 & 0.050 [0.022] & 0.048 [0.023] & 0.080 [0.024] & 0.066 [0.019]\\
19140 & 0.048 [0.021] & 0.050 [0.017] & 0.115 [0.040]  & 0.063 [0.020] \\
19830 & 0.166 [0.060] & 0.166 [0.054] & 0.563 [0.049] & 0.165 [0.069] \\
\hline
Overall Mean& 0.0835 & 0.0850 &  0.1623&0.0831\\
\hline
\end{tabular}
\label{tab:subjectwise_ksd}
\end{table}

\section{Discussion and conclusion}
This study investigated combining a point process model with RNN for beat-to-beat heartbeat dynamics modeling. 
The results demonstrate that RNN-based IGP parameter estimation is feasible and achieves performance comparable with results reported in previous study \cite{chen_characterizing_2010}.

However, the GOFs did not pass the KSD threshold for all subjects, indicating a mismatch between the IGP assumption and the observed R–R interval dynamics.
Several factors may contribute to this discrepancy. First, the IGP assumption may not fully capture the complexity of physiological heartbeat regulation, as previously suggested \cite{chen_characterizing_2010}.
Although recent approaches based on mixture lognormal formulations have shown better performance on this dataset \cite{subramanian2025density}, the present framework retains the IG model to preserve physiologically relevant parameters. Second, the current model only considers the R–R interval history, without incorporating additional physiological signals such as respiration or blood pressure, which could play an important role in heartbeat modulation. This suggests that extending the current framework to incorporate multi-modal physiological inputs may further improve model fit. Third, although extreme intervals were excluded, residual detection errors or ectopic beats may still be present, potentially violating the model assumptions.

As a conclusion, the proposed framework remains valuable as a method for tracking time-varying heartbeat dynamics directly from irregular beat-to-beat intervals. Future work will extend this framework to personalized heartbeat modeling and validate it in various conditions and populations with impaired autonomic regulation, e.g., diabetes patients.

\section*{Acknowledgments} 
Funded by the European Union (the HealthyW8 project). Views and opinions expressed are however those of the author(s) only and do not necessarily reflect those of the European Union or HaDEA. Neither the European Union nor the granting authority can be held responsible for them.

\bibliography{refs}
\begin{correspondence}
Ying Wang\\
Drienerlolaan 5, 7522 NB Enschede, The Netherlands\\
imwywk@gmail.com
\end{correspondence}

\end{document}